\documentstyle[11pt,newpasp,twoside]{article}
\def\edcomment#1{\iffalse\marginpar{\raggedright\sl#1\/}\else\relax\fi}
\reversemarginpar

\begin{document}
\title{THE FORMATION OF THE MILKY WAY DISK}
\author{Cristina Chiappini}
\affil{Department of Astronomy, Columbia University,
                 Mail Code 5247, Pupin Hall, 550 West 120th Street,
                 New York, NY 10027}
\author{Francesca Matteucci}
\affil{Dipartimento di Astronomia, Universit\`a di Trieste,
                 Via G.B. Tiepolo 11, I-34131 Trieste, Italy}
\author{Donatella Romano}
\affil{SISSA/ISAS, Via Beirut 2-4, I-34014 Trieste, Italy}

\begin{abstract}
We present theoretical results on the galactic abundance gradients of 
several chemical species for the Milky Way disk, 
obtained using an improved version of the two-infall model of 
Chiappini, Matteucci, \& Gratton (1997) that incorporates a more
realistic model of the galactic halo and disk. 
This improved model provides a satisfactory fit to the elemental
abundance gradients as inferred from the observations and also to other
radial features of our galaxy (i.e., gas, star formation rate and
star density profiles). We discuss the implications these results
may have for theories of the formation of the Milky Way and make some 
predictions that could in principle be tested by future observations.
\end{abstract}

\section{Results and Discussion}
In this work we adopt a chemical evolution model (see
Chiappini, Matteucci, \& Romano 2000) that assumes two 
main accretion episodes for the formation of the Galaxy: the first one 
forming the halo and bulge in a short timescale followed by a second one that 
forms the thin-disk, with a timescale which is an increasing function of the 
Galactocentric distance (being of the order of 7 Gyrs at the solar 
neighborhood). The present model takes into account in more detail than 
previously the halo density distribution and explores the effects of a 
threshold density in the star formation process, both during the halo and 
disk phases. The model also includes the most recent nucleosynthesis 
prescriptions concerning supernovae of all types, novae and single stars dying 
as white dwarfs. In the comparison between model predictions and available
data, we have focused our attention on abundance gradients as well as gas, 
star and star formation rate distributions along the disk, since this kind of 
model has already proven to be quite successful in reproducing the solar 
neighborhood characteristics. 

We suggest that the mechanism for the formation 
of the halo leaves heavy imprints on the chemical properties of the outer 
regions of the disk, whereas the evolution of the halo and the inner disk are 
almost completely disentangled. This is due to the fact that the halo and disk 
densities are comparable at large Galactocentric distances and therefore the 
gas lost from the halo can substantially contribute to build  up the outer 
disk. We also show that the existence of a threshold density for the star 
formation rate, both in the halo and disk phase, is necessary to reproduce 
the majority of observational data in the solar vicinity and in the whole 
disk. In particular, a threshold in the star formation implies the occurrence 
of a gap in the star formation at the halo-disk transition phase, in agreement 
with recent data.

Our main conclusions are:

\par\noindent{$\bullet$} Our best-model predicts gradients in good 
			 agreement with the observed ones in PNe, H\,{\small 
			 II} regions and open clusters.

\par\noindent{$\bullet$} The outer gradients are sensible to the halo 
			 evolution, in particular to the amount of halo 
gas which ends up into the disk. This result is not surprising since the 
halo density is comparable to that of the outer disk, whereas is negligible 
when compared to that of the inner disk. Therefore, the inner parts of the disk
($R$ $<$ $R_\odot$) evolve independently from the halo evolution.

\par\noindent{$\bullet$} We predict that the abundance gradients along the 
			 Galactic disk must have increased with time. 
This is a direct consequence of the assumed ``inside-out'' scenario for the 
formation of the Galactic disk. Moreover, the gradients of different elements 
are predicted to be slightly different, owing to their different nucleosynthesis 
histories. In particular, Fe and N, which are produced on longer timescales than 
the $\alpha$-elements, show steeper gradients. Unfortunately, the available 
observations cannot yet confirm or disprove this, because the predicted 
differences are below the limit of detectability.  

\par\noindent{$\bullet$} Our model guarantees  
                         a satisfactory fit not only to 
		         the elemental abundance gradients but it is also in 
			 good agreement with the observed radial profiles of 
			 the SFR, gas density and the number of stars in the 
			 disk.

\par\noindent{$\bullet$} Our best model suggests that the average 
$<[\alpha/$Fe]$>$ ratios in stars slightly decrease from 4 to 10 kpcs.
This is due to the predominance of disk over halo stars in this distance range 
and to the fact that the ``inside-out'' scenario for the disk predicts a decrease of 
such ratios. On the other hand we predict a substantial 
increase ($\sim 0.3$ dex) of these ratios in the range 10\,--\,18 kpcs, due to the 
predominance, in this region, of the halo over the disk stars.

Finally, we conclude that  a relatively short halo formation timescale ($\simeq$ 0.8 
Gyr), in agreement with recent estimates for the age differences among 
Galactic globular clusters, coupled with an ``inside-out'' formation of the 
Galactic disk, where the innermost regions are assumed to have formed much 
faster than the outermost ones, represents, at the moment, the most likely 
explanation for the formation of the Milky Way. This scenario allows us to 
predict abundance gradients and other radial properties of the Galactic disk 
in very good agreement with observations. 
More observations at large Galactocentric distances 
are needed to test our predictions.


\begin{references}
\reference
Chiappini, C., Matteucci, F., \& Gratton, R. 1997, \apj, 477, 765
\reference
Chiappini, C., Matteucci, F., \& Romano, D. 2000, \apj, (submitted)
\end{references}
\end{document}